\documentclass[aip,amsmath,amssymb,graphicx]{revtex4-1}

\draft % marks overfull lines with a black rule on the right
\usepackage{graphicx}% Include figure files
\usepackage{dcolumn}% Align table columns on decimal point
\usepackage{bm}% bold math
\usepackage[utf8]{inputenc}
\usepackage[T1]{fontenc}
\usepackage{mathptmx}
\usepackage{float}

\begin{document}

%\preprint{}

\title{Naked-Eye Ghost Imaging via Photoelectric-Feedback} %Title of paper



%\author{}
%\thanks{}
%\affiliation{}

\author{Gao. Wang}
\affiliation{ 
Electronic Materials Research Laboratory, Key Laboratory of the Ministry of Education \& International Center for Dielectric Research, School of Electronic and Information Engineering, Xi'an Jiaotong University, Xi'an 710049, China%\\This line break forced with \textbackslash\textbackslash
}%
 
\author{Huaibin. Zheng}%
 \email{huaibinzheng@xjtu.edu.cn}
\affiliation{ 
Electronic Materials Research Laboratory, Key Laboratory of the Ministry of Education \& International Center for Dielectric Research, School of Electronic and Information Engineering, Xi'an Jiaotong University, Xi'an 710049, China%\\This line break forced with \textbackslash\textbackslash
}%

\author{Yu. Zhou}
 %\homepage{http://www.Second.institution.edu/~Charlie.Author.}
\affiliation{%
MOE Key Laboratory for Nonequilibrium Synthesis and Modulation of Condensed Matter, Department of Applied Physics, Xi'an Jiaotong University, Xi'an 710049, China%\\This line break forced% with \\
}%

\author{Hui. Chen}
\affiliation{ 
Electronic Materials Research Laboratory, Key Laboratory of the Ministry of Education \& International Center for Dielectric Research, School of Electronic and Information Engineering, Xi'an Jiaotong University, Xi'an 710049, China%\\This line break forced with \textbackslash\textbackslash
}%

\author{Jianbin. Liu}
\affiliation{ 
Electronic Materials Research Laboratory, Key Laboratory of the Ministry of Education \& International Center for Dielectric Research, School of Electronic and Information Engineering, Xi'an Jiaotong University, Xi'an 710049, China%\\This line break forced with \textbackslash\textbackslash
}%

\author{Yuan. Yuan}
\affiliation{ 
Shaanxi Key Laboratory of Environment and Control for Flight Vehicle, Xi'an Jiaotong University, Xi'an 710049, China%\\This line break forced with \textbackslash\textbackslash
}%

\author{Fuli. Li}
 %\homepage{http://www.Second.institution.edu/~Charlie.Author.}
\affiliation{%
MOE Key Laboratory for Nonequilibrium Synthesis and Modulation of Condensed Matter, Department of Applied Physics, Xi'an Jiaotong University, Xi'an 710049, China%\\This line break forced% with \\
}%

\author{Zhuo. Xu}
\affiliation{ 
Electronic Materials Research Laboratory, Key Laboratory of the Ministry of Education \& International Center for Dielectric Research, School of Electronic and Information Engineering, Xi'an Jiaotong University, Xi'an 710049, China%\\This line break forced with \textbackslash\textbackslash
}%


\date{\today}

\begin{abstract}
Based on optical correlations, ghost imaging is usually reconstructed by computer algorithm from the acquired data. We here proposed an alternatively high contrast naked-eye ghost imaging scheme which avoids computer algorithm processing.	Instead, the proposed scheme uses a photoelectric feedback loop to realize the multiplication process of traditional ghost imaging. Meanwhile, it exploits the vision persistence effect to implement integral process and to generate negative images observed by naked eyes. To realize high contrast naked-eye ghost imaging, a special pattern-scanning architecture on a low-speed light-modulation mask is designed, which enables high-resolution imaging with lower-order Hadamard vectors and boosts the imaging speed as well. Moreover, two kinds of feedback circuits, the digital circuit and the analog circuit, are presented respectively, which can achieve high-speed feedback operation on the light intensity. With this approach, we demonstrate high-contrast real-time imaging for moving objects.
\end{abstract}

\pacs{}% insert suggested PACS numbers in braces on next line

\maketitle %\maketitle must follow title, authors, abstract and \pacs

\section{Introduction}
%\label{}
Since its inception in the 1990's ghost imaging (GI) has intrigued researchers due to its novel physical peculiarities and its potential possible applications. The typical ghost imaging setups consist of two correlated optical beams propagating in distinct paths and impinging on two spatially-separated photodetectors: the signal beam interacts with the object and then is received by a single-pixel (bucket) detector without spatial resolution, whereas the reference beam goes through an independent path and impinges on a spatial distribution detector, like charge-coupled device (CCD) without interacting with the object. Even though information from either one of the detectors used for the acquisition does not yield an image, an image can be obtained  by cross-correlating signals from bucket detector and CCD. 
The first GI, utilizing two-photon quantum entanglement, is reported by Pittman \textit{et\ al}\cite{RN9}. Later, it was demonstrated that GI could be implemented with pseudo-thermal sources\cite{RN5,RN6,RN3,RN2} and thermal light\cite{RN7}. In addition, the computational GI (CGI) with an improved setup is proposed by Shapiro\cite{RN8,RN10}, where the reference beam is instead by a computed field pattern. With the development of GI, this concept has been extended to domains beyond the usual optical domain mentioned above and outside of the capture of spatial proprieties of light. Recently, it  has been demonstrated with X-rays\cite{RN36,RN43,RN41,RN34}, atoms\cite{RN45}, and even electrons\cite{RN40} as well as temporal ghost imaging\cite{RN80,RN79,RN82,RN81,RN84,RN83,RN78}.

However, up to now, no matter what type of ghost imaging method is, the way to get the output image is usually reconstructed by computer algorithm from the acquired data. Here, we proposed an alternatively novel naked-eye ghost imaging scheme to avoid computer algorithm processing, which will promote GI's convenience. In detail, a photoelectric feedback loop is used to link the bucket detector and the light source, where the intensity of light source is modulated by each output current value of the bucket detector. That is to say the traditional GI's multiplication process between output current value of the bucket detector and corresponding value of intensity distribution of reference beam is realized by this new way of photoelectric feedback loop. It is important to recognize that there is inverse correlation in our work. Meanwhile, the vision persistence effect is used to implement integral process and to generate negative images observed by naked eyes.  In principle, all photosensitive material with the vision persistence effect can be competent for this integral imaging process to show the imaging result. 
To realize high contrast naked-eye ghost imaging, one of the challenges is overcoming the background introduced by the reference beam, since the image is immersed in the reference light beam. Toward this end, a special pattern-scanning architecture on a low-speed light-modulation mask is designed, which enables high-resolution imaging with lower-order Hadamard vectors and boosts the imaging speed as well. 
Moreover, two kinds of feedback circuits, the digital circuit and the analog circuit, are presented respectively, which can achieve high-speed feedback operation on the light intensity. With this approach, we demonstrate high-contrast real-time imaging for moving objects. Our work opens a new way to utilize GI and can be applied to those recently developed GI methods with the usual optical domain, X-rays, atoms and electrons or the field of LIDAR.

\section{Experimental Section}
\subsection{Experiment principle}
\begin{figure}[htb]
	\centering
	\includegraphics[width=0.95\linewidth]{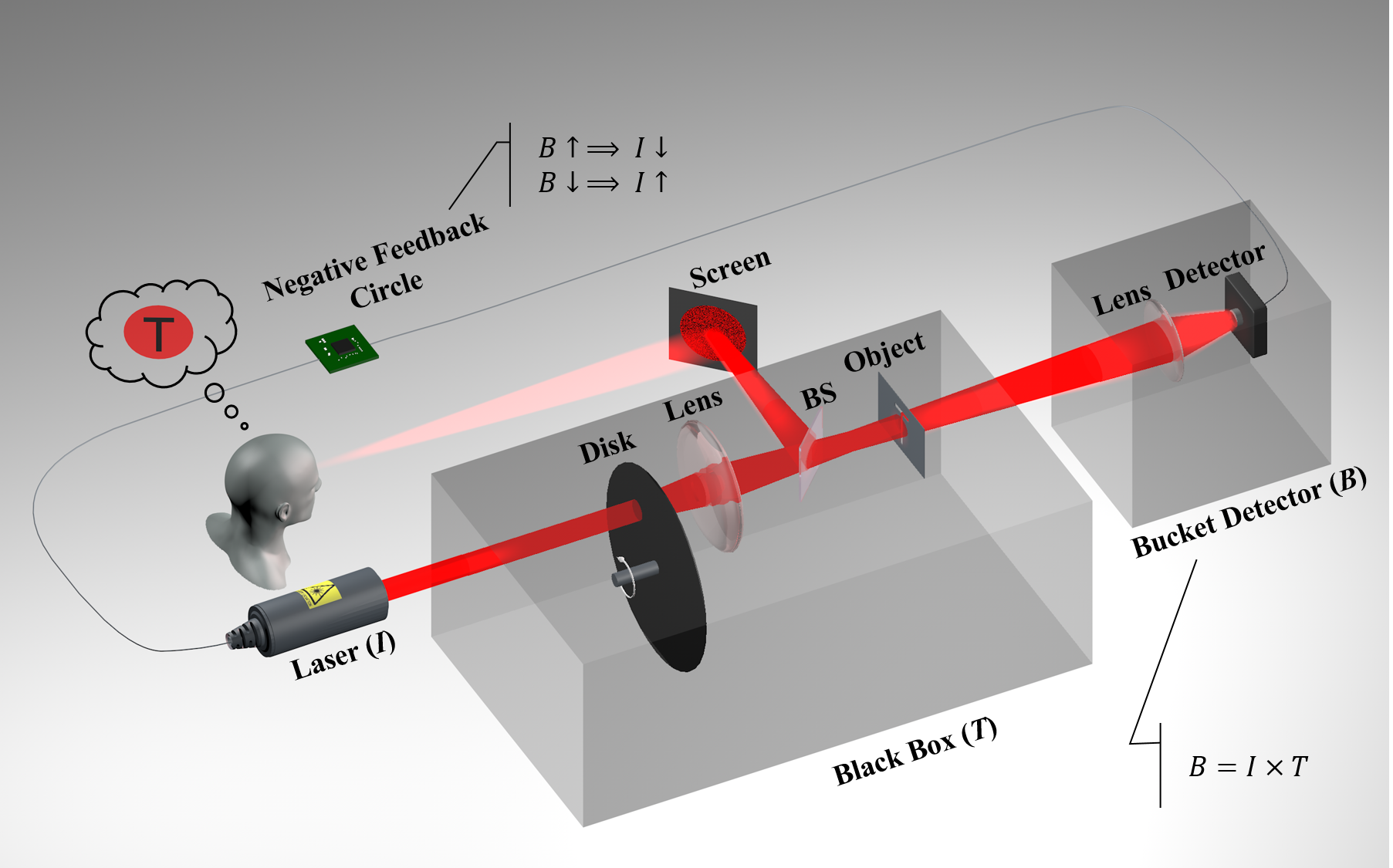}
	\caption{Schematic diagram of the naked-eye ghost imaging system, including a laser device, block box, bucket detector, feedback loop and naked-eye imaging. The block box with transmissivity $T_i$ is comprised of mask, lens, BS and object, respectively.}
	\label{fig:scheme}
\end{figure}

Figure \ref{fig:scheme} shows the schematic diagram of the naked-eye GI imaging system. One red laser beam is modulated by a rotating light-modulation mask. Then the modulated light is divided into two beams. The reflected beam is used to naked-eye imaging. The other illuminates and interacts with objects,  that are letters ``X'', ``J'', ``T'' and ``U'' with $ 35 \times 35 $ pixels, respectively. The transmitted light after objects is collected by a bucket detector comprising a collecting lens and a single-pixel photodetector. The output bucket signal is processed via the circuit and then becomes a feedback signal injecting into the laser driver, which modulates the laser intensity. So far, one loop for one pattern projection is completed. Once all loops for a group of patterns are completed in a time scale of vision persistence, a negative image will be observed by eyes at reflection arm mentioned above. 

In this work, we use a CCD camera to mimic the vision persistence effect of human eyes. Since the temporary retention time of human eyes is about 0.02 second in daytime vision, 0.1 second in intermediary vision and 0.2 second in night vision, we choose 0.2 second as the exposure time of CCD. At this point, a high-contrast real-time imaging will be observed by such photosensitive component.  

The differences from typical GI setup are that a photoelectric feedback loop is use to link the bucket detector and the light source, and the negative image can be observed directly by naked-eye at the position where the spatial distribution detector of typical GI is placed. To understand such naked-eye GI process, the imaging mechanism is shown in the following. 

Initially, the laser beam (with intensity ${I_i}$) goes though a black box (with transmissivity $T_i$) comprising a rotating mask and an object, and then is collected by a bucket detector as shown in Figure \ref{fig:scheme}. Thus, the output value ${B_i}$ of bucket detector is given by
\begin{equation}\label{B}
{B_i} = {I_i} \times {T_i}.
\end{equation}
Then, with this result, the algebraic loop can be built as follow 
\begin{equation}\label{I-t}
{I_{i + 1}} = f({B_i}) = f({I_i} \times {T_i}).
\end{equation}
From the aspect of statistics or steady-state, the Equation \ref{I-t} can be rewritten as 
\begin{equation}\label{I-I-K}
I = f(B) = f(I \times T).
\end{equation}

By introducing the negative feedback circuit, one can get that the modulated laser intensity $ I $ is a monotony decrease function of $ B $ value,
\begin{equation}\label{dfdB}
{\frac{df(B)}{dB}} < 0.
\end{equation}
Therefore, the laser intensity $ I $ has a significant inverse relationship with $ T $,
\begin{equation}\label{dIdK}
{{dI} \over {dT}} = {{df(B)} \over {dB}}{{dB} \over {dT}} = {{df(B)} \over {dB}}I < 0.
\end{equation}

%It is important to recognize that, in traditional GI system, $ T $ is the output value of the bucket detector and $ A $ is the pattern under the constant laser intensity. So, one can get the traditional ghost imaging result via the Equation \ref{G-traditional},

This system will degenerate into the traditional GI system without the feedback circle together with the intensity of light source be a constant value. Thus, the output value ${I_{2}(t_i)}$ of the bucket detector is only dependent on the transmissivity $T_i$ of the block box. Meantime, the intensity distribution ${I_{1}{(x,y,t_i)}}$ of the patterned light beam can be understood as the intensity of light source multiplying a mask modulation function  ${A_i}$. Therefore, images can be obtained via the correlation process\cite{RN96}, that is 
\begin{equation}\label{G-traditional}
G_{traditional}^{(2)} =  < {{I_{1}(x,y,t_i)}{I_{2}}(t_i)} >=< {A_i} {T_i} >,
\end{equation}
However, in our case, the intensity of light source is not constant, which is modulated by the feedback signal as shown in Equation \ref{I-I-K}. By substituting Equation \ref{I-I-K} into Equation \ref{G-traditional}, one can get naked-eye ghost imaging result via the Equation \ref{G-2},
\begin{equation}\label{G-2}
{\hat G^{(2)}} =  < {A_i}{I_i} >  =  < {P_i} >  = \int_t^{t + \tau} {{P_i}} dt,
\end{equation}
where the patterned light $ P_i $ is the result of that the feedback modulated laser beam interacts with the mask and goes through it, which realized the multiplication between $ A_i $ and $ I_i $. And then, this patterned light $ P_i $ is split by a beam splitter, where the transmission part illuminates the object serving as the feedback regulation and the reflected pattern is diffused on the screen. The output light from the screen is observed by a photosensitive component such as human eyes, performing integral imaging process, as shown in Equation \ref{G-2}, where $ \tau $ stands for the vision persistence time and ${\hat G^{(2)}}$ stands for the naked-eye ghost imaging result. Due to the inverse relationship, the negative image can be obtained. 

\subsubsection{Light modulation mask}

\begin{figure}[H]
	\centering
	\includegraphics[width=0.9\linewidth]{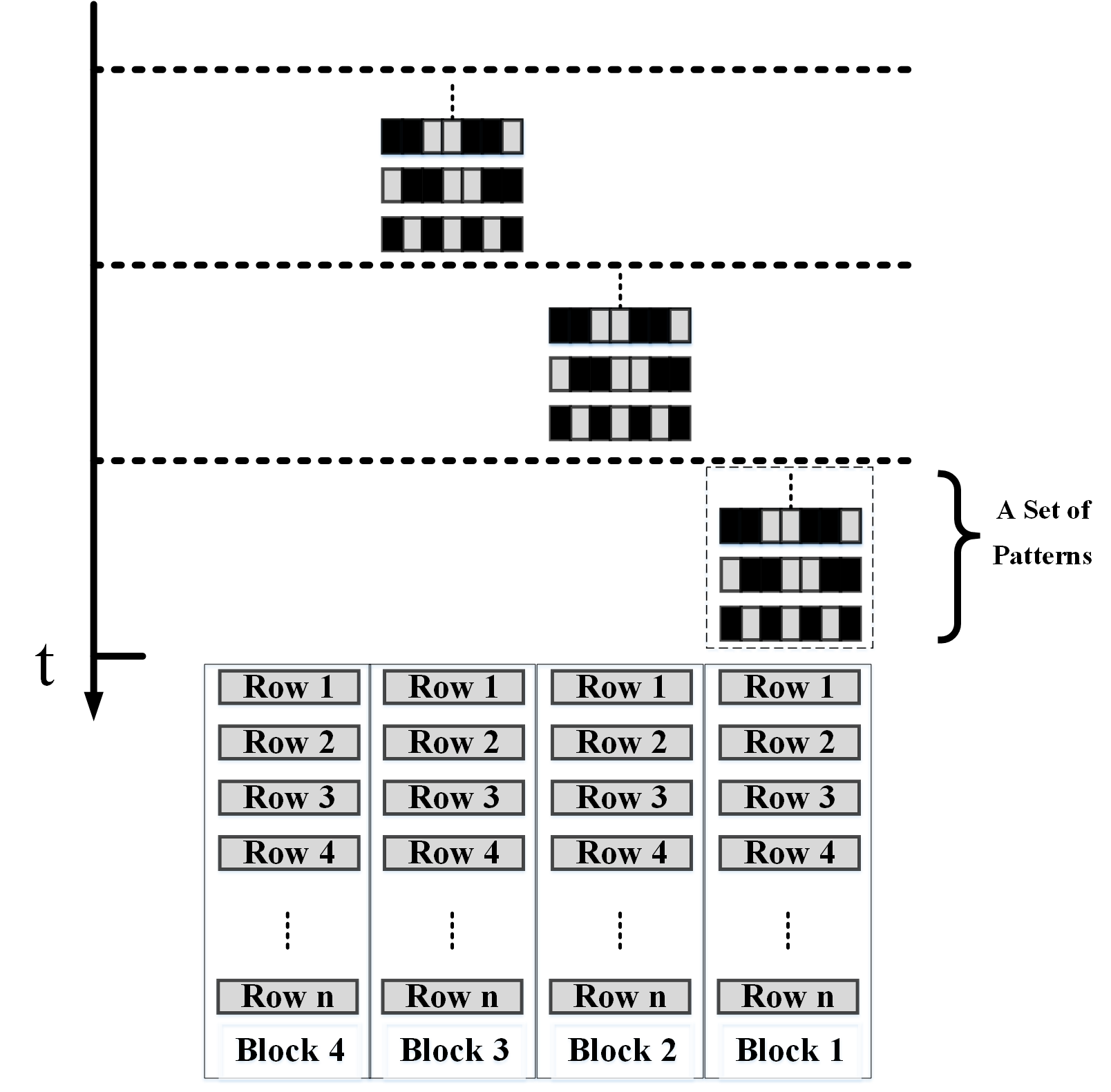}
	\caption{Structure of light modulation mask based on the Hadamard vector.}
	\label{fig:disk}
\end{figure}

One of the challenges with naked-eye GI is overcoming the background introduced by the reference beam, since the image is immersed in the reference light beam. To realize high contrast naked-eye GI, a special pattern-scanning architecture is designed on a low-speed light-modulation mask. 
Firstly, the object is divided into several blocks. Thus the dimensionality of the image can be reduced. For instance, one can divide the object ($ n\times n $) into $k$ column blocks. Every block is $ n\times (n/k) $ pixels.
Next, we use a complete set of low order Hadamard scanning pattern to scan each block row by row as shown in Figure \ref{fig:disk}.
So, one can get the visibility of each row in block via this imaging method,
\begin{equation}\label{contrast}
\begin{split}
Contras{t_{\hat H\_Block}} &= {{1 + {N_{Block}}} \over {1 + {N_{Block}}\left( {2{N_{Block}} - 5} \right)}},\\
{N_{Block}} &= n/k.
\end{split}
\end{equation}
In order to get high contrast via Hadamard pattern, apart from the sample scanning, it is a suitable choice that one can take ${N_{Block}} = 7$. In addition to high contrast property, this method enables high-resolution imaging with lower-order Hadamard vectors and boosts the imaging speed as well.

\subsection{Digital modulation method}
\begin{figure}[H]
	\centering
	\includegraphics[width=0.6\linewidth]{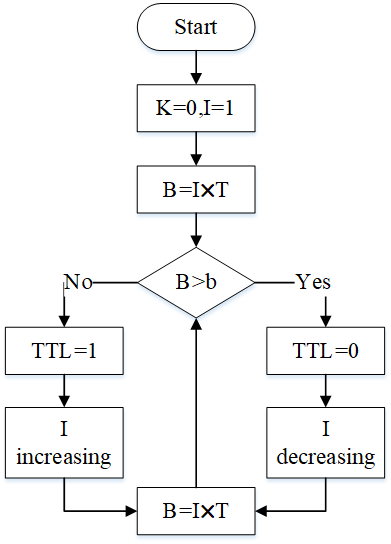}
	\caption{Work flowchart for the negative feedback digital modulation system.}
	\label{fig:DM}
\end{figure}

Figure \ref{fig:DM} shows the work flowchart for the negative feedback digital modulation system. The output signal $B$ of the bucket detector injects into this comparator to generate a TTL signal, which is fed into laser driver to modulate laser intensity. In detail, when $I> {b/T}$, $ I $ will decrease. When $I < {b/T}$, $ I $ will increase, where $b$ is the reference voltage. Taking the laser relaxation time into account, the laser intensity will approximate to
\begin{equation}\label{I-b-K}
I \approx {b \over T}.
\end{equation}
In addition, this imaging system formed by negative feedback has high control precision and stable operation, which can suppress noise apparently.

\subsection{Analog modulation method}
\begin{figure}[H]
	\centering
	\includegraphics[width=1\linewidth]{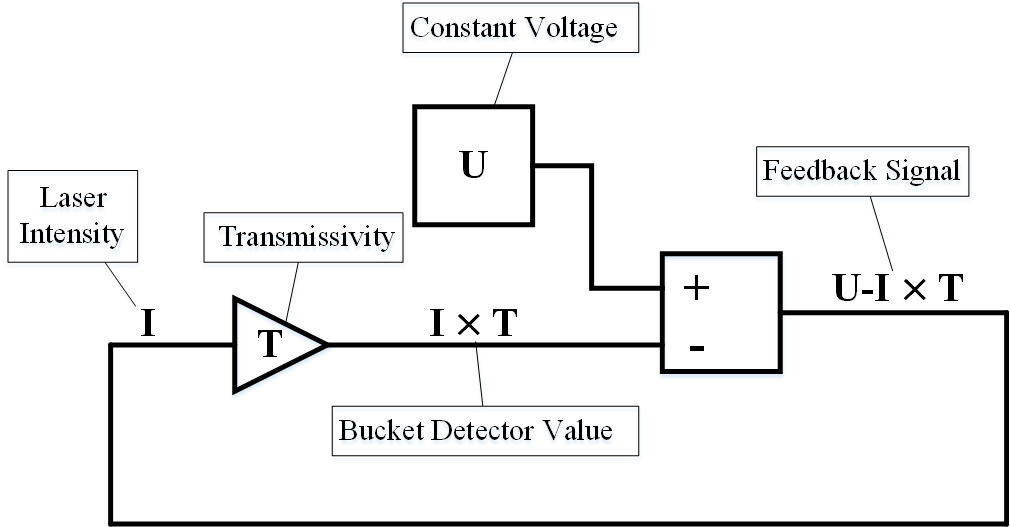}
	\caption{Negative feedback loop with an analog modulation.}
	\label{fig:AM}
\end{figure}

Experimental apparatus shown in Figure \ref{fig:scheme} can be re-described by a negative feedback loop when an analog modulation scheme is adopted, as shown in Figure \ref{fig:AM}. There are two input signals for analog modulator: one is a constant voltage ($ U $) denoting the measured laser intensity without any loss. The other is the output value $ B $ of bucket detector. The output signal ($ S $) from this analog modulator is the minus result between these two input signals 
  \begin{equation}\label{S}
S = I = U - I \times T.
\end{equation}
This signal is fed into the laser driver to modulate laser intensity. Considering the relaxation time of the imaging system, the laser intensity will be modulated by $\hat T$ as shown in Equation \ref{I-U-K}.
\begin{equation}\label{I-U-K}
\begin{split}
 I &= {U \over {\hat T}},\\
 {\hat T} &= 1 + T.
\end{split}
\end{equation}
The same effect such as noise suppression can be achieved by this system.

\subsection{Anti-noise capacity}
In this ghost imaging system, the main noise introduced is from the bucket detector, which is exposed to an external environment. 
\subsubsection{Anti-Noise on Circuits}
Firstly, in digital feedback loop, the noise signal from ambient noise light or others will be limit by the comparator of digital modulator. Moreover, if the noise goes through the comparator, the noise signal will be automatic suppression via the negative feedback system. 

Secondly, as well as the digital feedback loop, the analog feedback circle makes the output play the opposite role to the input of the noise, reducing the error between the system output and the system target. Ultimately, it makes the system tend to be stable.
\subsubsection{Anti-noise on imaging algorithm}
If the noise has been introduced into the system, the imaging expression will become 
\begin{equation}\label{G--1}
\begin{split}
G_I^{(2)} &= A_{N \times M}^T{({b \over {T - \Delta {T_{noise}}}})_{M \times 1}} \\
&= A_{N \times M}^T{({b \over T}{1 \over {1 - {{\Delta {T_{noise}}} \over T}}})_{M \times 1}}.
\end{split}
\end{equation}
Then one can get the Taylor expansion for Equation \ref{G--1}
\begin{equation}\label{G--2}
\begin{split}
G_I^{(2)} = A_{N \times M}^T({b \over T}(1 &+ ({{\Delta {T_{noise}}} \over T}) + {({{\Delta {T_{noise}}} \over T})^2}\\ + {({{\Delta {T_{noise}}} \over T})^3} 
&+ O{\left( {{{\Delta {T_{noise}}} \over T}} \right)^4}))_{M \times 1}.
\end{split}
\end{equation}
Because $\Delta {T_{noise}}$ is much small than $ T $, so it is easy to come out that
\begin{equation}\label{G--3}
{({{\Delta {T_{noise}}} \over T})^n} \ll {{\Delta {T_{noise}}} \over T} \ll 1,n \in {{\bf{N}}^*}.
\end{equation}
From the Equation \ref{G--2} and Equation \ref{G--3}, one can get that the effect of noise is significantly weakened.

\section{Results and discussion}

\begin{figure}[H]
	\centering
	\includegraphics[width=0.7\linewidth]{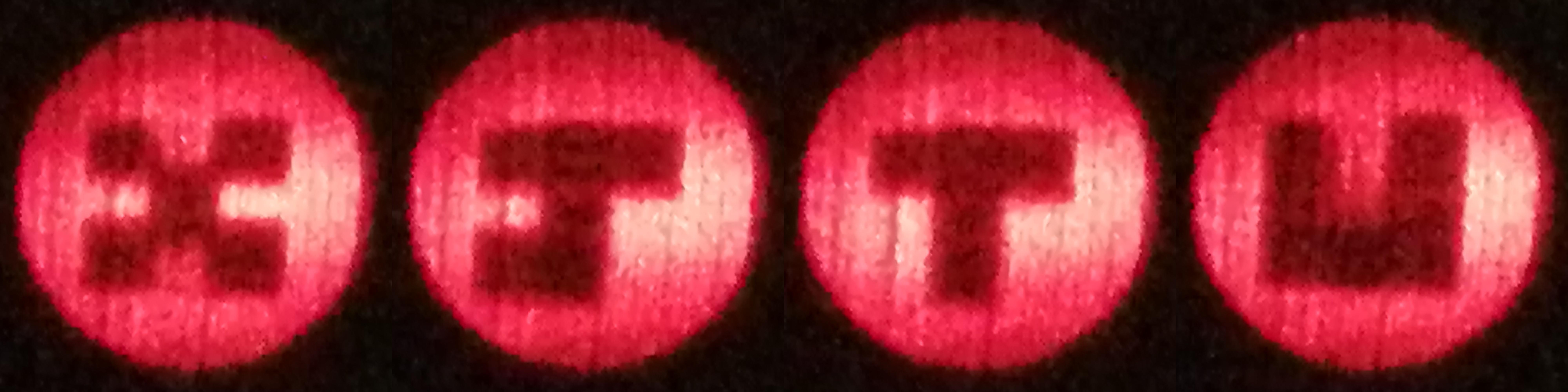}
	\caption{Imaging result under digital negative feedback.}
	\label{fig:xjtu1}
\end{figure}

\begin{figure}[H]
	\centering
	\includegraphics[width=0.7\linewidth]{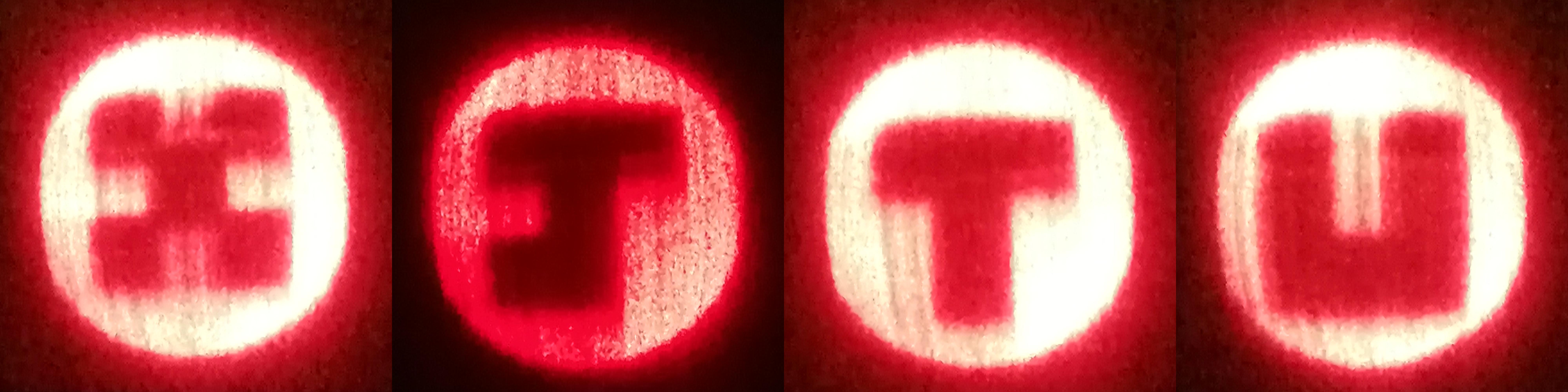}
	\caption{Imaging result under analog negative feedback.}
	\label{fig:xjtu2}
\end{figure}

Figure \ref{fig:xjtu1} and Figure \ref{fig:xjtu2} show the high contrast imaging results   (``X, J, T and U'', respectively) obtained by the digital negative feedback loop and analog negative feedback loop, respectively. Based on our method, the key problem of the image being immersed in the probe light beam is solved. It is worthy to note that they are negative images and videos.

From the imaging system, the modulator can be expressed as $ A \times X $, 
\begin{equation}\label{K-matrix}
{T_{M \times 1}} = A_{N \times M}^T{X_{N \times 1}}.
\end{equation}
where $ A $ is the mask modulation function and $ X $ denotes the object.
Equation \ref{B} can be rewritten via Equation \ref{K-matrix},
\begin{equation}\label{B-matrix}
{B_{M \times 1}} = diag({I_{M \times 1}})A_{N \times M}^T{X_{N \times 1}}
\end{equation}
On the other hand, through the Equation \ref{B}, $ T $ is equal to the bucket signal value when the laser intensity is constant. So, the traditional correlated imaging can get via $T $ and $ A $.

\begin{equation}\label{G-2-matrix}
{G^{(2)}} = A_{N \times M}^T{T_{M \times 1}}
\end{equation}
However the value of $ T $ is unknown, and only the laser intensity is changed with $ T $. In addition, from Equation \ref{I-b-K} and Equation \ref{I-U-K} , $ I $ and $ T $ are anti-related. 

As shown in Figure \ref{fig:scheme}, when one watches the screen, the human eye will automatically integrate the intensity modulated pattern. So, the imaging process can be express as 
\begin{equation}\label{key}
\begin{split}
G_{{I_1}}^{(2)} &= A_{N \times M}^T{I_{M \times 1}} = A_{N \times M}^T{({b \over T})_{M \times 1}},\\
G_{{I_2}}^{(2)} &= A_{N \times M}^T{I_{M \times 1}} = A_{N \times M}^T{({U \over {\hat T}})_{M \times 1}}.
\end{split}
\end{equation}

One can find that $ I $ and $ T $ are anti-related, resulting in negative images.
From the Equation \ref{I-b-K} and Equation \ref{I-U-K}, one can get the adaptive processing. If the pass speckle is very different from the object, the value of $T$ is very small. Thus the signal coming back is very weak and the bucket value $B$ is very small too. Thus, the system will automatically increase the intensity of the current speckle via the simple feedback system we proposed. Conversely, it will automatically decrease the intensity of the current speckle. As a result, the principal component of the negative image is strengthened.

\section{Conclusion}
In summary, a naked-eye ghost imaging via photoelectric feedback is realized. The obstacle to realizing high-contrast real-time imaging for moving objects is removed by a special pattern-scanning architecture and feedback system. Meanwhile, high resolution and the boosted imaging speed can be obtained with low pixel illumination from a low-speed rotating light-modulation mask. 
Two types feedback circuits, digital and analog, are used to modulate the laser intensity, which will bring the advantage of anti-noise. This work opens a new way to utilize GI, which has a potential application to 3D GI visualization, GI virtual reality and so on.

% If in two-column mode, this environment will change to single-column format so that long equations can be displayed. 
%\begin{widetext}
%$$\mbox{put long equation here}$$
%\end{widetext}

% Figures should be put into the text as floats. 
% Use the graphics or graphicx packages (distributed with LaTeX2e).
% Put the label that you will use with \ref{} command in the braces of the \label{} command.
%
% \begin{figure}
% \end{figure}

% \begin{table}
% \end{tabular}
% \end{table}

\begin{acknowledgments}
% Put your acknowledgments here.
National Basic Research Program of China (973 Program) (Grant No. 2015CB654602); Key Scientific and Technological Innovation Team of Shaanxi Province (Grant No. 2018TD-024); 111 Project of China (Grant No. B14040).\\
\end{acknowledgments}

\bibliography{VGI}

\end{document}